\documentclass{article}

\usepackage{PRIMEarxiv}

\usepackage{lineno,hyperref}
\usepackage{amsmath}
\usepackage{amsfonts}
\usepackage{amssymb}
\usepackage{geometry}
\usepackage{txfonts}
\usepackage{graphics}
\usepackage{graphicx}
\usepackage{bm}
\usepackage{booktabs}
\usepackage{multirow}
\usepackage{threeparttable}
\usepackage{caption}
\usepackage{subcaption}
\usepackage{arydshln}

\usepackage[utf8]{inputenc} 
\usepackage[T1]{fontenc}    
\usepackage{url}            
\usepackage{nicefrac}       
\usepackage{microtype}      
\usepackage{fancyhdr}       

\title{Wavelet neural operator: a neural operator for parametric partial differential equations}

\author{Tapas Tripura\\
  Department of Applied Mechanics\\
  Indian Institute of Technology Delhi\\
  \texttt{tapas.t@am.iitd.ac.in} \\
  \And
      Souvik Chakraborty \\
  Department of Applied Mechanics\\
  Yardi School of Artificial Intelligence (ScAI)\\
  Indian Institute of Technology Delhi\\
  \texttt{souvik@am.iitd.ac.in} \\
}

\begin{document}
\maketitle

\begin{abstract}
With massive advancements in sensor technologies and Internet-of-things (IoT), we now have access to terabytes of historical data; however, there is a lack of clarity in how to best exploit the data to predict future events. One possible alternative in this context is to utilize operator learning algorithm that directly learn nonlinear mapping between two functional spaces; this facilitates real-time prediction of naturally arising complex evolutionary dynamics. In this work, we introduce a novel operator learning algorithm referred to as the Wavelet Neural Operator (WNO) that blends integral kernel with wavelet transformation. WNO harnesses the superiority of the wavelets in time-frequency localization of the functions and enables accurate tracking of patterns in spatial domain and effective learning of the functional mappings. Since the wavelets are localized in both time/space and frequency, WNO can provide high spatial and frequency resolution. This offers learning of the finer details of the parametric dependencies in the solution for complex problems. The efficacy and robustness of the proposed WNO is illustrated on a wide array of problems involving Burger’s equation, Darcy flow, Navier-Stokes equation, Allen-Cahn equation, and Wave advection equation. Comparative study with respect to existing operator learning frameworks are presented. Finally, the proposed approach is used to build a digital twin capable of predicting Earth’s air temperature based on available historical data.
\end{abstract}

\keywords{Nonlinear mappings \and Operator learning \and Wavelet \and Wavelet Neural Operator \and Scientific machine learning.}

\section*{Introduction}
The application of partial differential equations (PDEs) for modelling of various physical and complex phenomena in scientific and engineering problems such as fluid flow, traffic flow, chemical reactions, biological growth etc \cite{renardy2006introduction,sommerfeld1949partial,jones2009differential}. are inevitable. In the absence of analytical solutions the large dimensional problems involving PDEs are traditionally solved using finite element \cite{hughes2012finite}, finite difference \cite{strikwerda2004finite} and finite volume \cite{eymard2000finite} based approaches. These traditional solvers are usually discretization dependent, due to which they are computationally expensive and requires independent forward runs for different parameter values in order to obtain the intended solutions. 
No need to mention the complexity of these traditional approaches as the dimension of the problem increases.In the recent years, neural networks (NN) have emerged as a possible alternative to the popular classical solvers. NNs approximate highly non-linear functions by passing the weighted sum of inputs through a non-linear activation function. NNs are universal approximators that can learn any functional mapping between two measurable space \cite{hornik1989multilayer}. 
The learning of NNs are either data-driven \cite{raissi2018deep,rudy2017data,wu2020data,navaneeth2022koopman} or physics-informed \cite{raissi2019physics,goswami2020transfer,chakraborty2021transfer} in nature. Physics-informed NNs are very accurate and ensures that the underlying governing law of the physical problem is satisfied. Therefore, if the physics of the underlying processes are known in advance then physics-informed learning of NN becomes practical. However, for most of the natural processes the governing physics are not know a-priory and training of NNs are done by resorting to data-driven approaches. One problem with the data-driven NNs is that they do not ensure the governing physics of the underlying problem, as a consequence they don not generalise beyond the training data. 
As an efficient and effective approach the trained networks should generalize for any unseen input function and predict the output with as much as low error. Towards addressing this challenges, recently, NNs are generalised to neural operators.

Neural operators learn the mapping between two infinite-dimensional function spaces and therefore, they are required to be trained only once. Once trained they can be used for prediction of the solution for any given input function. Similar to NNs, neural operators also learn the complex nonlinear operators by passing the global integral operators through the nonlinear local activation functions. Neural operators also share the same network parameters between different discretizations that makes them discretization invariant. Since, they are discretization invariant they provide huge computational advantages over classical PDE solvers. Similar to universal function approximation theory, the neural operators work on the theorem of universal operator approximation by a single-layer of neural networks \cite{chen1995universal}. 
Among the recently developed neural operators the DeepONet architecture was the first to learn such infinite-dimensional function spaces \cite{lu2019deeponet,lu2022comprehensive}. The DeepONet consists of two neural networks that are referred as branch and trunk NN. The branch NN is responsible for the input function and the trunk NN corresponds to the output function at some sensor point. The output in the DeeONet is  then obtained from the inner product of these NNs. Its application can be found in reliability analysis \cite{garg2022assessment}, bubble dynamics \cite{lin2021seamless}, fracture mechanics \cite{goswami2022physics} etc. 
In a different approach the graph neural operator (GNO) was proposed \cite{li2020neural}. The GNO focuses on learning of the infinite-dimensional mapping by composition of
nonlinear activation functions and certain class of integral operators. The kernel integration in the GNO is learned through message passing between the graph networks. Though development of GNO was completely different from classical NNs and gave new ideas of using graphs, the GNO has the tendency to become unstable with the increase of the number of hidden layers. In the same time, Fourier neural operator (FNO) was proposed with the aim to learn the parameters of the network in Fourier space \cite{li2020fourier}. The heart of the FNO is the spectral decomposition of the input using fast Fourier transform (FFT) and computation of the convolution integral kernel in the Fourier space. Simple yet accurate and efficient performance of the FNO was demonstrated on various state-of-the art problems including Burgers, Darcy and Navier-Stokes equations. 

One major shortcoming of the FNO is that the basis functions in FFTs are generally frequency localised with no spatial resolution \cite{bachman2000fourier}. Therefore the FFTs are not suitable for studying the spatial behaviour of any signal or image, as a consequence, the performance of the FNO gets hindered in complex boundary conditions. In this context, one can think of wavelets which are both space and frequency localized \cite{bachman2000fourier,boggess2015first}. Since wavelets have spatial information they can handle signal with discontinuity and/or spikes better and therefore can learn the patterns in images better compared to FFTs.
To explain further, wavelet provides us characteristic frequency and characteristic space information, with the help of which we can not only tell the frequencies present in the signal as a function of characteristic frequency but also the location of the frequency as a function of characteristic space information. 
The application of wavelets can be found in various aspects of data compression such as fingerprint compression \cite{wirsing2020time}, compressed sensing \cite{mallat1992singularity}, iris recognition \cite{cho2002method}, signal denoising \cite{sheybani2011dimensionality}, motion analysis \cite{martin2011novel}, fault detection \cite{liu2012shannon}, to name few. Use of wavelets in neural networks can also be found in the literature \cite{zhang1995wavelet,xu2019graph,shervani2020chemical,gupta2021multiwavelet}.

In this work, we propose a new spectral decomposition based neural operator for learning mappings between infinite-dimensional spaces of functions and call this Wavelet Neural Operator (WNO). In the proposed WNO, the snapshots are first transformed to its high and low frequency components using discrete wavelet transform \cite{selesnick2005dual,ray2006dual}. In general, the high frequency components represent the edges in an image and the low frequency components account for the smooth regions. Since in a convolution setup the prime aim is to detect the edges present in a image, the high frequency components are of more importance than the low frequency components in our case. Thus, the single tree discrete wavelet transform with multi-level discrete wavelet transform of the high frequency components are utilized in this work. 
The proposed WNO harnesses the benefits of the wavelets and can learn solution operators of highly nonlinear parametric PDEs with both smooth and discontinuity either on the boundary conditions or within the domain. It can further be shown that, if the mother wavelet basis in WNO is replaced by trigonometric functions then FNO becomes the special case of WNO. Overall, the salient points of the proposed WNO architecture are as follows:
\begin{itemize}
    \item The network parameters in the proposed WNO are learned in the wavelet space that are both frequency and spatial localized, thereby can learn the patterns in the images and/or signals more effectively.
    \item The proposed WNO can handle highly nonlinear family of PDEs with discontinuities and abrupt changes in the solution domain and the boundary.
    \item The accuracy of the proposed WNO is consistent for both smooth and complex geometry.
    \item The proposed WNO can learn the frequency of changes in the patterns over the solution domain, which makes it amenable for online implantation's without requiring for the entire dataset. Thus proposed WNO can be generalised to images and videos, as a consequence proposed WNO can also be implemented for short-to-medium range climate modelling and weather forecast purposes.
\end{itemize}
Due to these reasons, WNO can be straightforwardly applied to a wide array of problems including fluid mechanics, traffic flow modelling, and climate modelling. The efficacy and robustness of the proposed WNO is illustrated using several examples involving Burger's equation, Darcy flow, Navier-Stokes equation, and Wave advection equation. Comparative study with respect to other existing neural operators has been presented. Finally, we utilize the proposed WNO to develop a digital twin capable of predicting Earth's air temperature at 2m above the ground at a resolution of $2^\circ \times 2^\circ$ based on historical data.

\section*{Results}
\subsection*{Operator learning through neural operator}\label{sec:methods}
An operator is responsible for mapping between infinite-dimensional input and output function spaces. To understand it better, let $D \in \mathbb{R}^{d}$ be the $n$-dimensional domain with boundary $\partial D$. For a fixed domain $D=(a,b)$ and $x \in D$, consider a PDE which maps the input function spaces i.e., the space containing the source term $f(x, t) : D \mapsto \mathbb{R}$, the initial condition $u(x,0) : D \mapsto \mathbb{R}$, and the boundary conditions $u(\partial D, t) : D \mapsto \mathbb{R}$ to the solution space $u(x, t) : D \mapsto \mathbb{R}$, with $t$ being the time coordinate.
In the present work, we focus on learning the operator that maps the input functions to the solution space $u(x,t)$. Now, let us define two complete normed vector spaces $(\mathcal{A}, \|.\|_{\mathcal{A}})$ and $(\mathcal{U}, \|.\|_{\mathcal{U}})$ of functions taking values in $\mathbb{R}^{d_a}$ and $\mathbb{R}^{d_u}$ with given number of partial derivatives. 
These function spaces are often called as Banach spaces and denoted as $\mathcal{A}:= \mathcal{C}(D;\mathbb{R}^{d_w})$ and $\mathcal{U}:= \mathcal{C}(D;\mathcal{R}^{d_u})$. For the given domain $D \subset \mathbb{R}^{d}$ and normed spaces $(\mathcal{A}, \|.\|_{\mathcal{A}})$ and $(\mathcal{U}, \|.\|_{\mathcal{U}})$, we aim at learning the nonlinear differential operator $\mathcal{D}: \mathcal{A} \mapsto \mathcal{U}$. 
The modern data-driven and physics-informed NNs center around the learning of the function $u(x, t)$ that satisfies the differential operator $\mathcal{D}$ in the domain $D$ and boundary conditions $\partial D$. Although these approaches are cost effective than classical PDE solvers, they can still be computationally expensive due to the fact that they need to be trained multiple times for multiple set of PDE parameters ${\bm{a}} \in \mathcal{A}$. One elegant approach is to learn the nonlinear differential operator $\mathcal{D}$ so that the solutions to a family of PDEs can be obtained for different sets of input parameters ${\bm{a}} \in \mathcal{A}$, which results in computationally efficiency and has more practical utility. 

\begin{figure}[ht!]
	\centering
	\includegraphics[width=\textwidth]{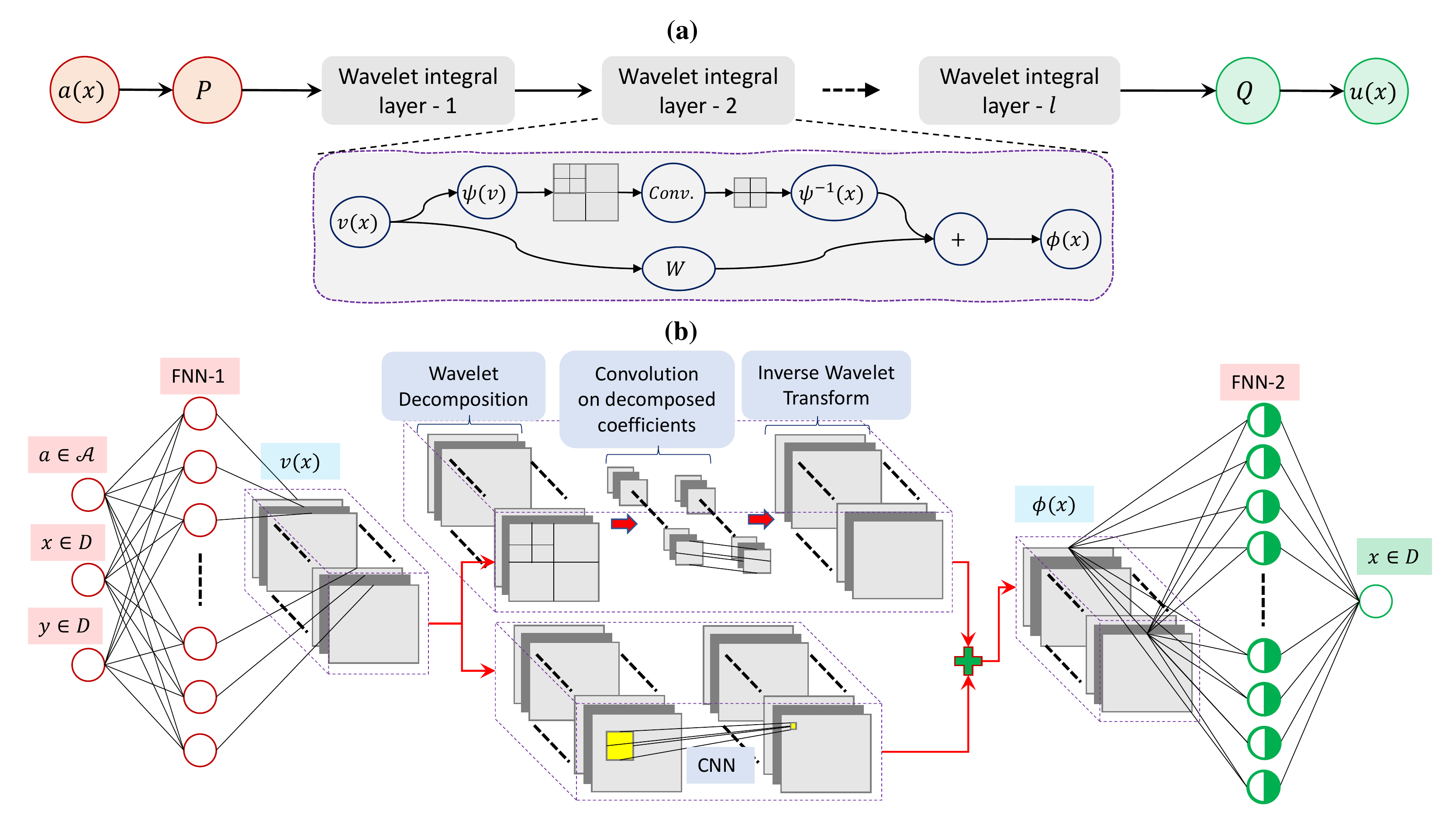}
	\caption{\textbf{Illustration of the architecture of the wavelet neural operator (WNO)}. (a) \textbf{Schematic of the proposed neural operator}. First lift the inputs to a higher dimension by a local transformation $P(\cdot)$. Then pass the lifted inputs to a series of wavelet kernel integral layer. Transform back the final output of the wavelet integral layer using a local transformation $Q(\cdot)$. Finally, activate the output of $Q(\cdot)$, which will provide the solution $u(x)$. In the wavelet kernel integral layer, the multilevel wavelet decomposition of the inputs are performed first, as a result of which the horizontal, vertical and diagonal coefficients at different levels are obtained. The convolution between neural network weights and the subband coefficients of last level are then performed. The inverse wavelet transform is performed on the convoluted coefficients to transform the reduced inputs to original dimension. Suitable activation is done on the output of the wavelet kernel integration layer. (b) \textbf{A simple WNO with one wavelet kernel integral layer}. The inputs contain the initial parameters and space information. The local transformations $P(\cdot)$ and $Q(\cdot)$ are modelled as shallow fully connected neural network. The output of $P(\cdot)$ is fed into the wavelet integral layer. The integral layer consists of two separate branches. In the first branch the wavelet decomposition of inputs followed by parameterization of integral kernel is done. In the second branch a convolution neural network (CNN) with kernel size 1 is constructed. The outputs of the two branches are then summed and activation's are performed. Then the outputs are passed through the transformation $Q(\cdot)$, which provides the target solution $u(x)$. In similar manner, a WNO with arbitrary number of wavelet integral layers can be be constructed.}
	\label{fig_methodology}
\end{figure}

Let, the input and output functions in $\mathcal{A}$ and $\mathcal{U}$ are denoted by ${\bm{a}} : D \mapsto a(x \in D) \in \mathbb{R}^{d_a}$ and ${\bm{u}}: D \mapsto u(x \in D) \in \mathbb{R}^{d_u}$. Further, consider that $u_j = \mathcal{D}(a_j)$ and we have access to $N$ number of pairs $\{(a_j, u_j)\}_{j=1}^N$. Then, we aim to approximate $\mathcal{D}$ by neural networks from the pairs $\{(a_j, u_j)\}_{j=1}^N$ as,

\begin{equation}
    \mathcal{D} : \mathcal{A} \times {\bm{\theta}}_{NN} \mapsto \mathcal{U}; \quad {\bm{\theta}_{NN}} = \left\{ \mathcal{W}_{NN}, {\bm{b}}_{NN} \right\},
\end{equation}
where ${\bm{\theta}_{NN}}$ denotes the finite-dimensional parameter space of the neural networks. 
Although the input and output functions $a_{j}$ and $u_{j}$ are continuous, however, in the data-driven framework we assume the $n$ point discretization $\{x_p\}_{p=1}^n$ of the domain $D$. Therefore, for the $N$ collection of the input-output pair we assume that we have access to the datasets $\{ a_{j} \in \mathbb{R}^{n \times d_{a}}, u_{j} \in \mathbb{R}^{n \times d_{u}} \}_{j=1}^N$. With this setup, the objective here is to develop a network that can exploit the data to learn the operator $\mathcal D$. To that end, 
we first lift the inputs $a(x,t) \in \mathcal{A}$ to some high dimensional space through a local transformation $P(a(x)): \mathbb{R}^{d_a} \mapsto \mathbb{R}^{d_v}$ and denote the high dimensional space as $v_{0}(x)=P(a(x))$, where $v_{0}(x)$ takes values in $\mathbb{R}^{d_v}$. In a neural network framework, the local transformation $P(a(x))$ can be achieved by constructing a shallow fully connected neural network (FNN). Further, let us denote that a total number of $l$ steps are required to achieve satisfactory convergence.
Therefore, we apply $l$ numbers of updates $v_{j+1} = G(v_{j}) \forall j=1,\ldots,l$ on $v_{0}(x)$, where the function $G: \mathbb{R}^{d_v} \mapsto \mathcal{R}^{d_v}$ takes value in $\mathbb{R}^{d_v}$. 
Once all the updates are performed, we define another local transformation $Q(v_{l}(x)): \mathbb{R}^{d_v} \mapsto \mathbb{R}^{d_u}$ and transform back the output $v_{l}(x)$ to the solution space $u(x)= Q\left(v_{l}(x)\right)$.
The step-wise updates $v_{j+1} = G(v_{j})$ are defined as, 

\begin{equation}\label{eq:iteration}
    v_{j+1}(x):= g \left( \left(K(a; \phi) * v_{j}\right)(x) + W v_{j}(x) \right); \quad x \in D, \quad j \in [1,l]
\end{equation}
where $g(\cdot): \mathbb{R} \to \mathbb{R}$ is a non-linear activation function, $W: \mathbb{R}^{d_{v}} \to \mathbb{R}^{d_{v}}$ is a linear transformation, $K: \mathcal{A} \times {\bm{\theta}} \to \mathcal{L}\left(\mathcal{U}, \mathcal{U}\right)$ is the integral operator on $\mathcal{C}\left(D ; \mathbb{R}^{d_{v}}\right)$.
In the neural network setup, we represent $K(a ; \phi)$ to be a kernel integral operator parameterized by $\phi \in {\bm{\theta}}$. The kernel integral operator $K(a ; \phi)$ is defined as,

\begin{equation}\label{eq:integral}
    \left(K(a ; \phi) * v_{j}\right)(x):=\int_{D \in \mathbb{R}^{d}} k \left(a(x,y), x, y; \phi \right) v_{j}(y) \mathrm{d} y; \quad x \in D, \quad j \in [1,l]
\end{equation}
where $k_{\phi}: \mathbb{R}^{2d+d_{a}} \mapsto \mathbb{R}^{d_{v} \times d_{v}}$ is a neural network parameterized by $\phi \in {\bm{\theta}}$. Suppose, there is some kernel $k \in \mathcal{C}\left(D; \mathbb{R}^{d_v} \right)$, then we want to employ the concept of degenerate kernels to learn the $\kappa_{\phi}$ from the data such that, 

\begin{equation}\label{eq:converge}
    \| \left( k_{\phi}(a) - k(a) \right) v \| \le \| \left( k_{\phi}(a) - k(a) \right) \| \|v \|,
\end{equation}
and $k_{\phi} \to k$. Since the kernel $k \in \mathcal{C}(D)$ and $K \in \mathcal{L}(\mathcal{C}(D;\mathbb{R}^{d_v}))$ together defines the infinite-dimensional spaces, therefore, Eq. \eqref{eq:iteration} can learn the mapping of any infinite-dimensional function space. As far as the approximation of nonlinear operators are concerned the structure of Eq. \eqref{eq:iteration} follows the conventional neural networks. We propose a new neural operator referred to as the WNO that learns the mapping with higher order of accuracy. One key component of the proposed WNO is the wavelet kernel integral layer.
The overarching framework of the proposed WNO is presented in Fig. \ref{fig_methodology}. While Fig. \ref{fig_methodology}(a) provides the schematic description of proposed WNO, in Fig. \ref{fig_methodology}(b), a simplistic WNO with wavelet kernel integral layer is shown for easy understanding.

For training the network, we define an appropriate loss function $\mathcal{L} = \mathcal{U} \times \mathcal{U}$ as follows:

\begin{equation}
    {\bm{\theta}}_{NN} = \mathop {\arg \min}\limits_{{\bm{\theta}}_{NN}} \mathcal{L} \left(\mathcal{D}({\bm{a}}), \mathcal{D}(\bm{a}, \bm{\theta}) \right).
\end{equation}
We leverage the universal approximation theorems of the neural networks \cite{hornik1989multilayer,chen1995universal} and try to learn the parameter space (weights-$\mathcal{W}_{NN}$ and biases-$\bm{b}_{NN}$ of neural networks) by approaching to the loss function from a data-driven perspective. 

\subsection*{Operator learning of example problems through WNO}\label{sec:numerical}
Numerical case studies on a wide variety of systems representing various class of problems in fluid dynamics, gas dynamics, traffic flow, atmospheric science and phase field modelling are considered in this section. In all the case, the performance of the network architectures were compared using the $L^2$ relative error of the predictions. We compare the results obtained with four popular neural operators and their variants: (a) DeepONet \cite{lu2022comprehensive,lu2019deeponet}, (b) Graph Neural Operator (GNO), (c) Fourier Neural Operator (FNO) \cite{li2020fourier}, and (d) Multiwavelet Transformation operator (MWT) \cite{gupta2021multiwavelet}.
The overall WNO architecture consists of four layers of WNO, each of which are activated by the GeLU activation function \cite{hendrycks2016gaussian}. 
The mother wavelet is chosen from the Daubechies family \cite{daubechies1992ten,meyer1993wavelets}. For all the cases, the sizes of the individual datasets, the wavelet, the level of wavelet tree and the number of layers in the network against each problem are listed in the Table. \ref{tab:network_param}. For optimization of the parameters of the WNO, the ADAM optimizer with initial learning rate of 0.001 and a weight decay of $10^{-6}$ is utilized. During the optimization, decay in the learning rate of each parameter group are introduced at every 50 epochs at a rate of 0.75. The total number of epochs used for training the WNO architecture against the undertaken examples are given in Table \ref{table_epochs}. The batch size in the data loader varies according to the underlying system between 10-25. The numerical experiments are performed on a single RTX A2000 6GB GPU. A summary of results for different example problems solved are shown in Table \ref{table_accuracy}. It is observed that the error for the proposed WNO varied in the range of 0.21\% -- 1.75\% with WNO yielding the best results (lowest error) in four out of the six cases. In the remaining two problems also, the results obtained using the proposed WNO are highly accurate (and close to the best results).


\begin{table}[ht!]
    \centering
    \caption{Dataset size for each problem, unless otherwise stated. WNO architecture for each problem, unless otherwise stated.}
    \label{tab:network_param}
    \begin{tabular}{p{4cm}lllllllll}
    \hline
    \multirow{2}{*}{Examples} & \multicolumn{2}{c}{Number of data} & \multirow{2}{*}{Wavelet} & \multirow{2}{*}{m} & \multicolumn{4}{c}{Network dimensions} & \multirow{2}{*}{$g(.)$} \\ \cline{2-3}\cline{6-9}
     & Training & Testing & & & FNN1 & FNN2 & CNN & WNO & \\
    \hline
    Burgers & 1000 & 100 & db6 & 8 & 64 & 128 & 4 & 4 & GeLU\\
    Darcy (rectangular) & 1000 & 100 & db4 & 4 & 64 & 128 & 4 & 4 & GeLU\\
    Darcy (triangular) & 1900 & 100 & db6 & 3 & 64 & 128 & 4 & 4 & GeLU \\
    Navier-Stokes & 1000 & 100 & db4 & 3 & 26 & 128 & 4 & 4 & GeLU\\
    Allen-Cahn & 1400 & 100 & db4 & 1 & 64 & 128 & 4 & 4 & GeLU\\
    Advection & 900 & 100 & db6 & 3 & 96 & 128 & 4 & 4 & GeLU\\
    \hline
    \end{tabular}
\end{table}

\begin{table}[ht]
	\centering
	\caption{Number of epochs required to obtain the best prediction results. Note that the number of epochs here for DeepONet and FNO, except the Allen-Cahn example are reported from their respective original papers.}
	\label{table_epochs}
	\begin{tabular}{p{3cm} llllll}
		\hline
		& \multicolumn{6}{c}{Number of epochs} \\ \cline{2-7}
		Architectures & Burgers' & Darcy flow & Navier-Stokes & Allen-Cahn & Darcy (Notch) & Advection \\
		\hline
		DeepONet & 500000 &  100000 & 100000 & 100000 & 20000 & 250000 \\
		POD-DeepONet & 500000 &  100000 & 100000 & 100000 & 20000 & 250000 \\
		FNO & 500 & 500 & 500 & 500 & 500 & 500 \\
		MWT & 500 & 500 & 500 & 500 & 500 & 500 \\ \hdashline
		WNO & 500 & 500 & 500 & 500 & 500 & 500 \\
		\hline
	\end{tabular}
\end{table}

\begin{figure}[ht!]
	\centering
	\begin{subfigure}[b]{0.7\textwidth}
         \centering
         \includegraphics[width=\textwidth]{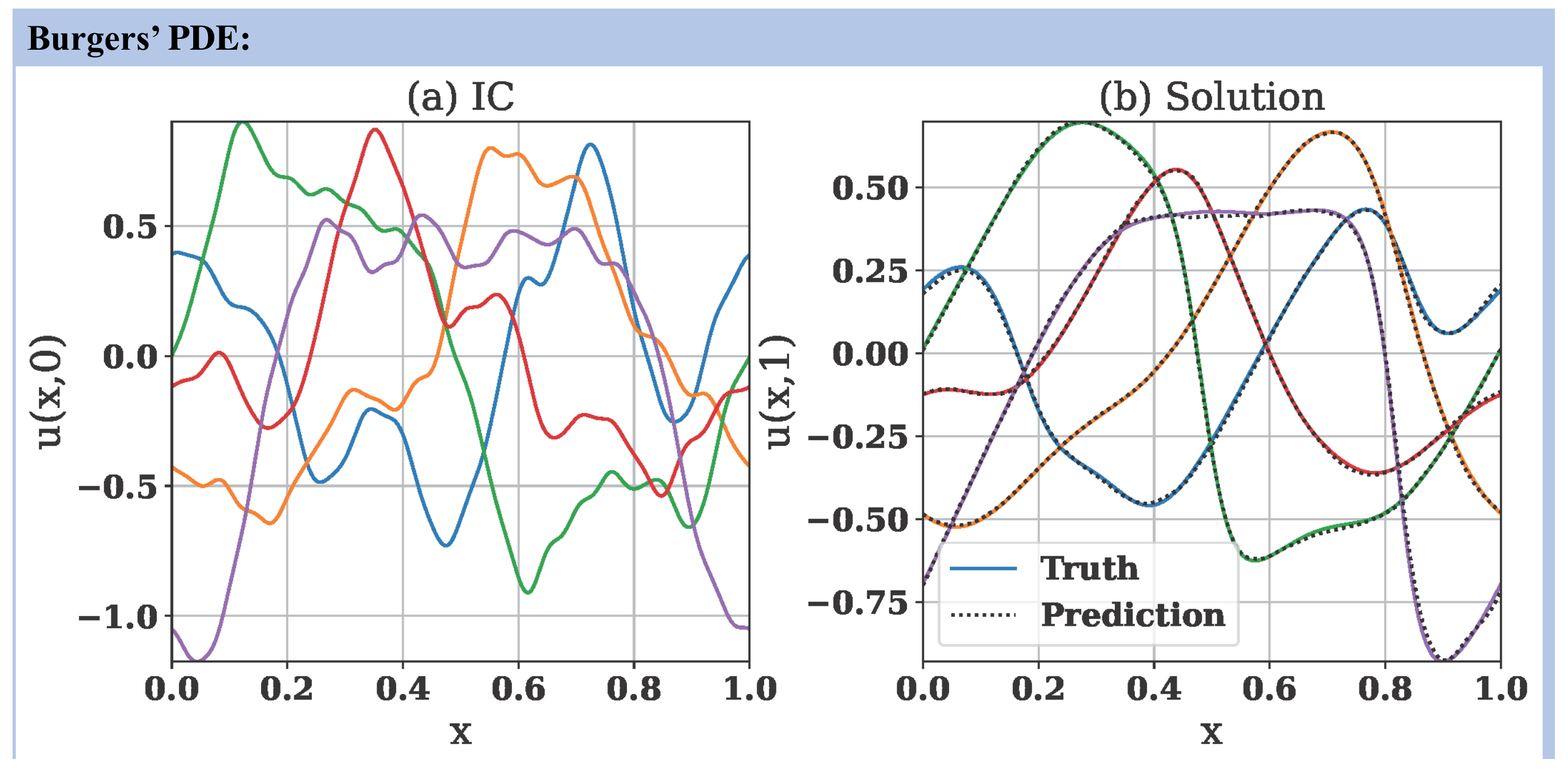}
     \end{subfigure}
     \begin{subfigure}[b]{\textwidth}
         \centering
         \includegraphics[width=\textwidth]{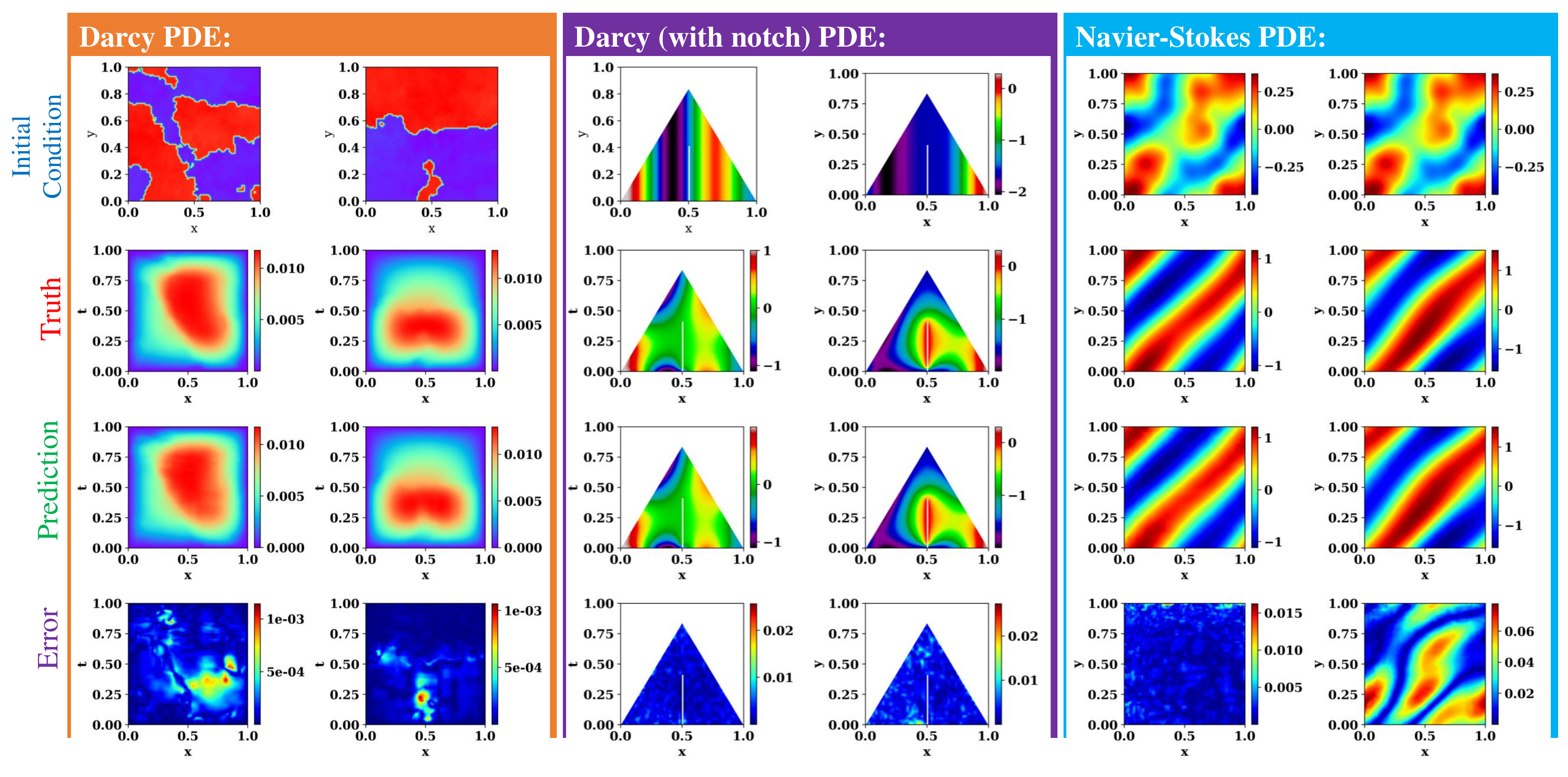}
     \end{subfigure}
	\caption{\textbf{Operator learning of flow problems}. First row: \textbf{Burgers' PDE with 1024 spatial resolution}. (a) the initial conditions. (b) The solution to the 1D Burgers flow equation at $t=1$. The aim was to learn the differential operator of Burgers' PDE. For given input functions, the predictions are shown and it can be seen that the predictions of proposed WNO matches with the true solution almost exactly. \textbf{Darcy flow in the rectangular domain with spatial resolution of ${\bf{85 \times 85}}$}. The aim is to learn the differential operator of Darcy equation. The prediction of output pressure fields from two different input permeability conditions are shown in the figure. The errors indicate that the predictions match the true solutions almost exactly. \textbf{Darcy flow in a triangular domain with a notch}. The goal here is to test the performance of proposed WNO for learning nonlinear operators in complex geometric conditions. For the given boundary conditions, the pressure fields obtained using the proposed WNO are shown in the figure. The errors against each of the predictions depict that the proposed WNO can learn operators in complex geometric conditions also. \textbf{Navier-Stokes equation with spatial resolution of ${\bf{64 \times 64}}$}. In this example, the aim is to use WNO for learning of differential operators of 2D time dependent problems. The initial vorticity (top row) and the predictions at $t=11$s (second row first column) and $t=20$s (second row second column) are shown. The errors indicates that the proposed WNO can learn highly nonlinear operators.}
	\label{fig_result1}
\end{figure}

\textbf{1D Burgers equation}: In the first example, we consider the 1D Burgers equation. The 1D Burgers equation is popularly used in modelling of 1D flows in fluid mechanics, gas dynamics, and traffic flow. The 1D Burgers equation is given by the form,
$\partial_{t} u(x, t)+\frac{1}{2}\partial_{x}u^{2}(x, t) =\nu \partial_{x x} u(x, t); x \in(0,1), t \in(0,1]$, subjected to the initial condition $u(x, 0) = u_{0}(x); x \in(0,1)$, where, $\nu \in \mathbb{R}^{+*}$ is the viscosity of the flow and $u_{0}(x) \in L^2_{per}\left((0,1); \mathbb{R} \right)$. A periodic boundary condition of the form, $u(x - \pi, t) = u(x + \pi, t); x \in(0,1), t \in(0,1]$ is considered.
The aim here is to learn the operator, $\mathcal{D}: u_{0}(x) \mapsto u(x,1)$. The datasets are taken from the Ref. \cite{li2020fourier}, where $\nu$=0.1 and a spatial resolution of 1024 is considered. 
\textbf{Results}: the prediction results along with their accuracy are illustrated in the Fig. \ref{fig_result1} and Table \ref{table_accuracy}. 
From the error values in the Table. \ref{table_accuracy}, it is evident that the MWT obtains the lowest error followed by FNO. The proposed WNO has slightly higher error than FNO but lesser than DeepONet and POD-DeepONet. Although from the quantitative standpoint the differences in the error in the predictions using WNO is slightly higher than MWT and FNO, from Fig. \ref{fig_result1}, the differences in the predictions and true results are not discerned. This indicates that, although not the best, the proposed WNO performs reasonably well for this problem.

\begin{table}[ht]
	\centering
	\caption{Mean $L^2$ relative error between the true and predicted results.}
	\label{table_accuracy}
	\begin{threeparttable}
	\begin{tabular}{p{3.5cm} llllll:l}
		\hline
		\multirow{2}{*}{PDE} & \multicolumn{7}{c}{Network Architectures}\\ \cline{2-8}
		& GNO & DeepONet & FNO & MWT & POD-DeepONet\tnote{\ddag} & dgFNO+\tnote{$\dagger$} & WNO \\
		\hline
		Burgers' equation & $\approx$ 6.15 \% & $\approx$ 2.15 \% & $\approx$ 1.60 \% & $\approx$ 0.19 \%  & $\approx$ 1.94 \% & - & $\approx$ 1.75 \% \\
		Darcy (rectangular) & $\approx$ 3.46 \% & $\approx$ 2.98 \% & $\approx$ 1.08 \% & $\approx$ 0.89 \%  & $\approx$ 2.38\% & - & $\approx$ 0.84 \% \\
		Darcy (triangular) & - & $\approx$ 2.64 \% & - & $\approx$ 0.87 \%  & $\approx$ 1.00 \% & $\approx$ 7.82\% & $\approx$ 0.77 \% \\
		Navier-Stokes equation & - & $\approx$ 1.78 \% & $\approx$ 1.28 \% & $\approx$ 0.63 \%  & $\approx$ 1.71 \% & - & $\approx$ 0.31 \% \\
		Allen-Cahn & - & $\approx$ 17.7 \% & $\approx$ 0.93 \% & $\approx$ 4.84 \%  & - & - & $\approx$ 0.21 \% \\
		Wave-advection & - & $\approx$ 0.32 \% & $\approx$ 47.7 \% & $\approx$ 10.22 \%  & $\approx$ 0.40 \% & $\approx$ 0.62 \% & $\approx$ 0.62 \% \\
		\hline
	\end{tabular}
	\begin{tablenotes}
	\item $\ddag$ POD-DeepONet is the modified version of DeepONet, proposed in Ref. \cite{lu2022comprehensive}. $\dagger$ Note that in complex geometric condition FNO does not work, thus we use dgFNO+, which is a modified version of FNO (Ref. \cite{lu2022comprehensive}).
	\end{tablenotes}
	\end{threeparttable}
\end{table}

\textbf{2D Darcy flow equation in a rectangular domain}: The 2D Darcy flow equation is widely used in modelling of flow, whether liquid or gas, through a  porous medium. In the absence of the time component it represents a second order nonlinear elliptic PDE and the form in a 2D domain is given as,
$-\nabla \cdot \left(a(x,y) \nabla u(x,y) \right) =f(x,y); x,y \in(0,\mathbb{R})$, subjected to the initial condition, $u(x,y) = u_0({x,y}); x,y \in \partial(0,\mathbb{R})$, where $a(x,y)$ is the permeability field, $u(x,y)$ is the pressure, $\nabla u(x,y)$ is the pressure gradient, and $f(x,y)$ is a source function. In this problem definition, the Darcy flow is defined on a unit box domain with $x \times y \in (0, 1)^2$ with $u_0(x,y) = 0$ (zero Dirichlet boundary condition). The aim is to learn the operator, $\mathcal{D}: a(x,y) \mapsto u(x,y)$. The datasets for training the network architectures are taken from the Ref. \cite{li2020fourier}. During training a spatial resolution of 85 $\times$ 85 is used. 
\textbf{Results}: the predictions for the 2D Darcy flow in rectangular domain is plotted in Fig. \ref{fig_result1}, while the associated errors are given in Table. \ref{table_accuracy}. In this case, it is observed that the mean prediction error for WNO is lowest among all the considered methods. It is further evident in the Fig. \ref{fig_result1}, where the differences between true and predicted solutions can not be discerned in naked eye.

\textbf{2D Darcy flow equation with a notch in triangular domain }:
This example emulates the previous 2D Darcy problem but with more complex boundary condition. The setup and datasets for this case are taken from the Ref. \cite{lu2022comprehensive}, where the boundary conditions for the triangular domain are generated using the following Gaussian process (GP),
$u(x) \sim {\rm{GP}}(0, \mathcal{K}(x,x'))$. The kernel is taken as $\mathcal{K}(x,x') = \exp \left(-{(x-x')^2}/{2l^2} \right); l=0.2; x,x' \in [0,1]$.
In addition to the triangular geometry, a notch is introduced in the flow medium. The permeability field $a(x,y)$ and the forcing function $f(x,y)$ are set as 0.1 and -1, respectively. The aim here is to learn the operator, responsible for the mapping of the boundary conditions to the pressure field in the entire domain, represented as, $\mathcal{D}: u(x,y)|_{\partial \omega} \mapsto u(x,y)$. 
\textbf{Results}: the testing results with corresponding testing errors are given in Fig. \ref{fig_result1} and Table \ref{fig_result1}, respectively. From the error values in table it can be observed that our proposed WNO has the lowest prediction error, which is more than three fold lower than DeepONet and ten fold less than FNO. The errors between the truth and prediction can also be observed in Fig. \ref{fig_result1}, that shows that the WNO is able to map the given boundary conditions to the pressure filed correctly. Upon careful observation, it can further be noticed that the predictions are not only correct in the smooth regions but also provides near perfect match near the slit. This indicates that the proposed WNO can be applied for effective and accurate learning of nonlinear operators in complex geometric conditions.

\textbf{2D time dependent Navier-Stokes equation}: The Navier-Stokes equation is a second order nonlinear parabolic PDE and its application can be found in various facets of fluid flow problems such as air flow around the aeroplane wing, ocean currents, thermodynamics, etc. The 2D incompressible Navier-Stokes equation in the vorticity-velocity form is given by the equations, $\partial_{t} \omega(x,y, t)+u(x,y, t) \cdot \nabla \omega(x,y, t) =\nu \Delta \omega(x,y, t)+f(x,y); x, y \in(0,1), t \in(0, T]$ and $\nabla \cdot u(x,y, t) =0; x, y \in(0,1), t \in[0, T]$, where $\nu \in \mathbb{R}$ and $f(x,y)$ are the viscosity of the fluid and source function. The initial vorticity field is taken as $\omega(x,y, 0) = \omega_{0}(x,y); x, y \in(0,1)$. The terms $u(x, y, t)$ and $\omega(x, y, t)$ are the velocity and vorticity field of the fluid, respectively. The viscosity in our experiment is considered as $\nu = 10^{-3}$. The aim here is to learn the operator responsible for the mapping between the vorticity field at the time steps $t \in [0, 10]$ to the vorticity at a target time step $t \in (10,T]$ i.e., $\omega |_{(0,1)^2 \times [0,10]} \mapsto \omega |_{(0,1)^2 \times (10,T]}$. The initial vorticity $\omega_0{(x,y)}$ is simulated from a Gaussian random field. For this example, the datasets for initial vorticity $\omega_0{(x,y)}$ and the solution $\omega_{10}{(x,y)}$ at time $t=10$ are taken from Ref. \cite{li2020fourier}. For both training and testing, the spatial resolution of the vorticity fields are fixed at resolutions 64 $\times$ 64. 
\textbf{Results}: the prediction results for vorticity field at $t \in \{10,20\}$ are given in Fig. \ref{fig_result1}, along with the mean prediction errors in Table \ref{table_accuracy}. The results show that the prediction results of the proposed WNO has the lowest error and emulates the true solutions almost accurately.

\begin{figure}[ht!]
	\centering
	\includegraphics[width=\textwidth]{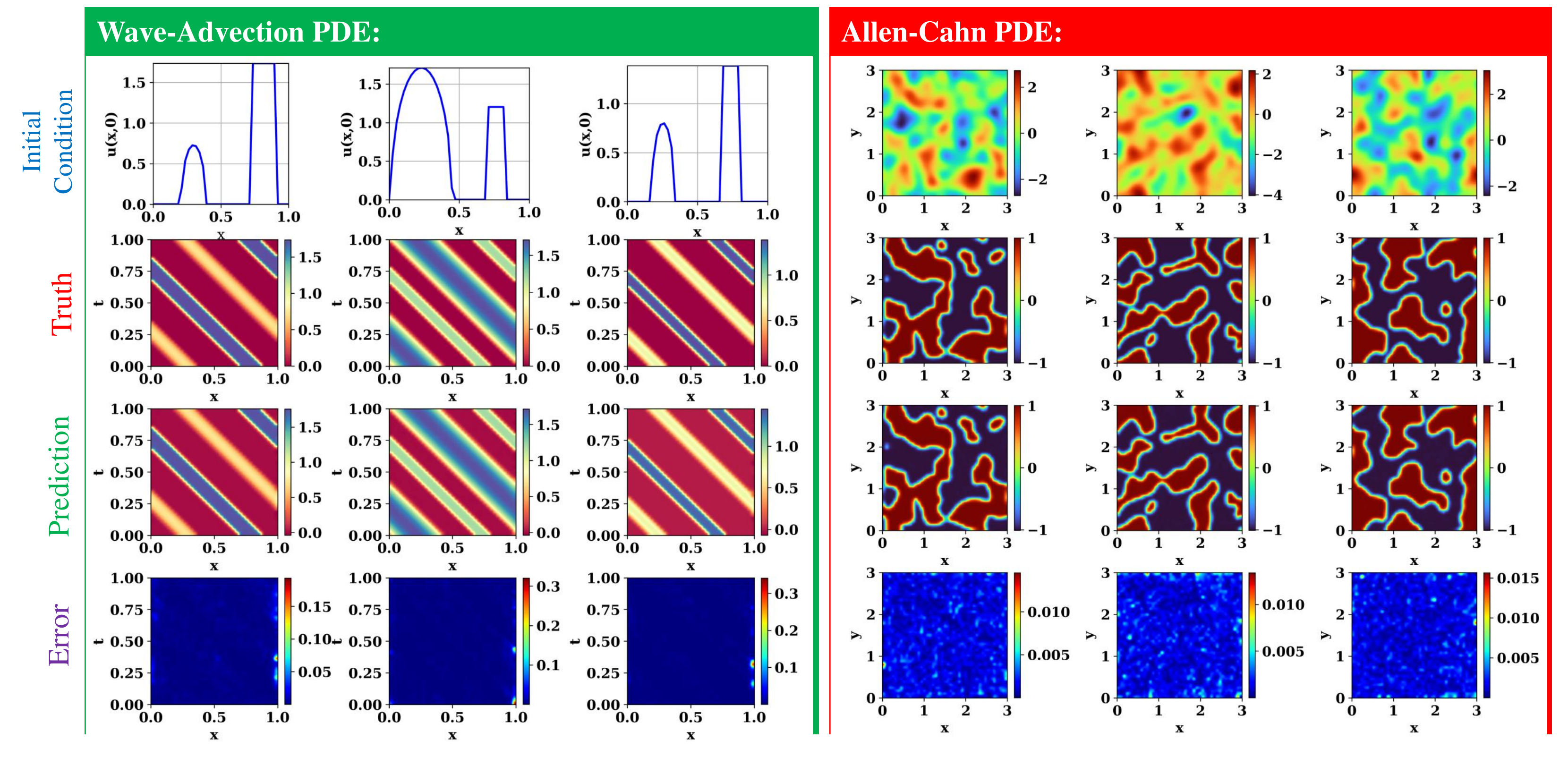}
	\caption{\textbf{Operator learning of Allen-Cahn reactor-diffusion equation and wave advection equation}. \textbf{Wave advection PDE}: For solution, a spatio-temporal resolution of $40 \times 40$ is used. The proposed WNO is used here to learn the operator that can predict the wave propagation through continuous medium. Three independent initial conditions (IC) were taken, for which the true, prediction and corresponding errors are plotted. The error plots demonstrate that the prediction results of proposed WNO is very accurate. \textbf{Allen-Cahn PDE}: For three different initial conditions, the final solutions at $t=20$s are plotted. This PDE has chaotic nature, i.e., the solutions diverges significantly for small perturbations in initial conditions. The aim here is to test the ability of the proposed WNO for prediction of chaotic type solutions. The solutions are obtained on a spatial resolution of $43 \times 43$. Three different IC are taken and for each IC the true and WNO predicted solutions are shown. The errors against each case demonstrate that the proposed WNO could predict the true solutions almost exactly.}
	\label{fig_acandav}
\end{figure}

\textbf{2D Allen-Cahn equation}: The Allen–Cahn equation is a reaction–diffusion equation and often used in the context of chemical reactions and modelling of phase separation in multi-component alloy. The Allen-Cahn equation in 2-dimensional space is given as,
$\partial_t u(x,y,t) = \epsilon \Delta u(x,y,t) + u(x,y,t) - u(x,y,t)^3; x,y \in (0,3)$, subjected to the initial condition $u(x,y,0) = u_0(x,y); x,y \in (0,3)$, where $\epsilon \in \mathbb{R}^{+*}$ is a real positive constant and responsible for the amount of diffusion. The problem is defined on periodic boundary with $\epsilon = 10^{-3}$ and the initial condition is simulated from the Gaussian Random Field using the following kernel, $\mathcal{K}(x,y)= \tau^{(\alpha-1)}\left(\pi^2(x^2+y^2) + \tau^2 \right)^{\alpha/2}$ with $\tau=15$ and $\alpha=1$. 
The aim here is to learn the operator $\mathcal{D}:u_{0}(x,y) \mapsto u(x,y,t)$. In this case, $t$=20s is taken and the solution is obtained on a grid of 43 $\times$ 43. 
\textbf{Results}: The prediction results with mean prediction error of the proposed WNO for different input functions are given in Fig. \ref{fig_acandav} and Table \ref{table_accuracy}. 
The mean errors in Table \ref{table_accuracy} indicates that the DeepONet and MWT has significant error. The FNO has shown good performance with $\approx 1 \%$ error, however, our proposed WNO obtains the predictions with lowest error which is closes to $0.2 \%$. In Fig. \ref{fig_acandav}, it is straightforward to see that the true and prediction results are non-differentiable for all the given input functions. This further illustrates that the proposed WNO can be successfully implemented for any 2D time dependent highly nonlinear problems. 

\textbf{Wave advection equation}: The wave advection equation is a hyperbolic PDE and primarily describes the solution of a scalar under some known velocity field. The advection equation with periodic boundary condition is given by the expression, $\partial_t u(x,t) + \nu \partial_x u(x,t) = 0; x \in (0,1), t\in (0,1)$ and $u(x - \pi, t) = u(x+\pi, t); x \in (0,1), t \in (0,1)$.
Here $\nu \in \mathbb{R} > 0$ represents the speed of the flow. The initial condition is chosen as, $u(x,0) = h1_{ \left\{c- \frac{\omega}{2}, c+\frac{\omega}{2} \right\} } + \sqrt{ \max(h^2 - (a(x-c))^2,0) }$.
where the variables $\omega$ and $h$ represents the width and height of the square wave, respectively and the wave is centered at $x = c$. The values of $\{c, \omega, h\}$ are randomly chosen from $[0.3, 0.7] \times [0.3, 0.6] \times [1, 2]$. For $\nu$=1, the solution to the advection equation is given as, $u(x,t)=u_0(x-t)$. The aim here is to learn the operator, that defines the mapping $\mathbb{G}: u_0(x) \mapsto u(x,t)$, for some later time $t$. The solutions are obtained on a spatial resolution of 40 $\times$ 40. The datasets for this example are taken from the Ref. \cite{lu2022comprehensive}. 
\textbf{Results}: The prediction results along with their $L^2$ relative error are given in Fig. \ref{fig_acandav} and Table \ref{table_accuracy}. Similar to the previous examples, it can be seen in Fig. \ref{fig_acandav} that the WNO is able to predict the scalar field for arbitrary input function. The corresponding errors show that the FNO and MWT suffers the most with $\approx47\%$ and $\approx 10\%$ error, respectively. The DeepONet yields the lowest error, however, both POD-DeepONet, modified FNO and proposed WNO also obtains relatively small error (very close to the DeepONet results).

\begin{figure}[ht!]
	\centering
	\includegraphics[width=\textwidth]{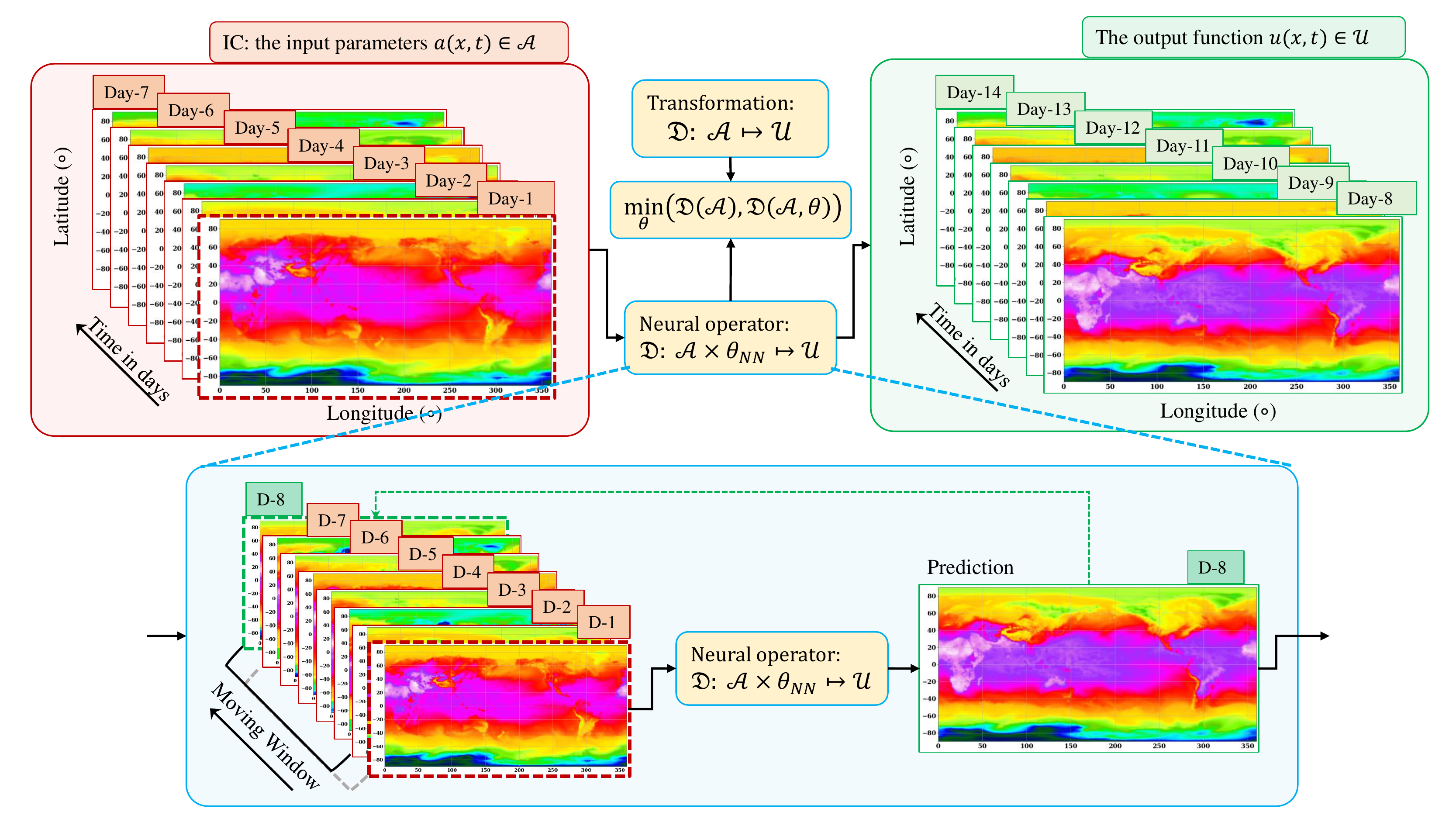}
	\caption{\textbf{Methodology for the short-term weekly forecast of 2m air temperature on a resolution of ${\mathbf{2^{\circ} \times 2^{\circ}}}$}. The aim of the short-term forecast is to train the WNO using previous 7 days data and then predict the 2m temperature for next 7 days. To do this, a recurrent neural network (RNN) structure is employed within the WNO. The RNN framework takes the previous 7 days data as input and then performs prediction for Day 8. For prediction the parameters are learned through convolution using the proposed WNO architecture. The Day 8 prediction data is then appended in the previous 7 day database. To maintain the recurrent structure i.e., to transfer the information from Day 8 to future predictions the information of Day 1 from previous data is discarded and the process is continued until the next 7 day predictions are made. This results in a highly efficient and accurate.}
	\label{fig_era5_climate}
\end{figure}

\subsection*{Weather forecast: prediction of the monthly mean 2m air-temperature}\label{sec:application}
The previous examples have showcased the effectiveness of the proposed WNO framework in accurately solving parametric PDEs. In general sense, however, the solutions of the PDEs can be imagined as videos of snapshots of the time-evolving process. Since, the images and videos presents the state of an object at a particular time and their motions, respectively, they can be idealised as 2D and 2D time dependent problems. Thus, the applications of WNO can also be extended for learning of dynamical systems from images and videos recorded using phone, high-speed camera or satellites. Further, once the WNO is trained, the predictions stage is extremely efficient as compared to traditional methods; this makes WNO an ideal candidate when it comes to real-time prediction. In this section,
we implement the proposed WNO to develop a digital twin that can predict Earth's temperature. In particular, we use WNO for accurate short to medium-range prediction of the hourly and mean monthly 2m temperature of air based on historical data. The datasets are taken from European Centre for Medium-Range Weather Forecasts (ECMWF) which provides publicly available hourly and monthly averaged datasets (so called ERA5) of various parameters of the climate. The ERA5 datasets are the fifth generation ECMWF reanalysis of various parameters of the atmosphere that uses data assimilation techniques to combine the numerical weather prediction data with freshly obtained observations to create new improved predictions of the various states of the atmosphere. The ERA5 database is a huge collection of hourly and monthly measurements for a large number of atmospheric, ocean-wave and land-surface parameters. The hourly and monthly datasets are stored both on single levels (atmospheric, ocean-wave and land surface quantities) and different pressure levels (upper air fields). The resolution of the datasets are maintained as $0.25^{\circ} \times 0.25^{\circ}$ for atmosphere, $0.5^{\circ} \times 0.5^{\circ}$ for ocean waves and $0.1^{\circ} \times 0.1^{\circ}$ for land. While there are many variables present and many combinations are possible for various pressure levels, we only focus on one parameter with a spatial resolution of $2^{\circ} \times 2^{\circ}$. Combining all the parameters to develop a complete digital twin for Earth's climate is left as a future research direction.
\begin{figure}[ht!]
	\centering
	\includegraphics[width=\textwidth]{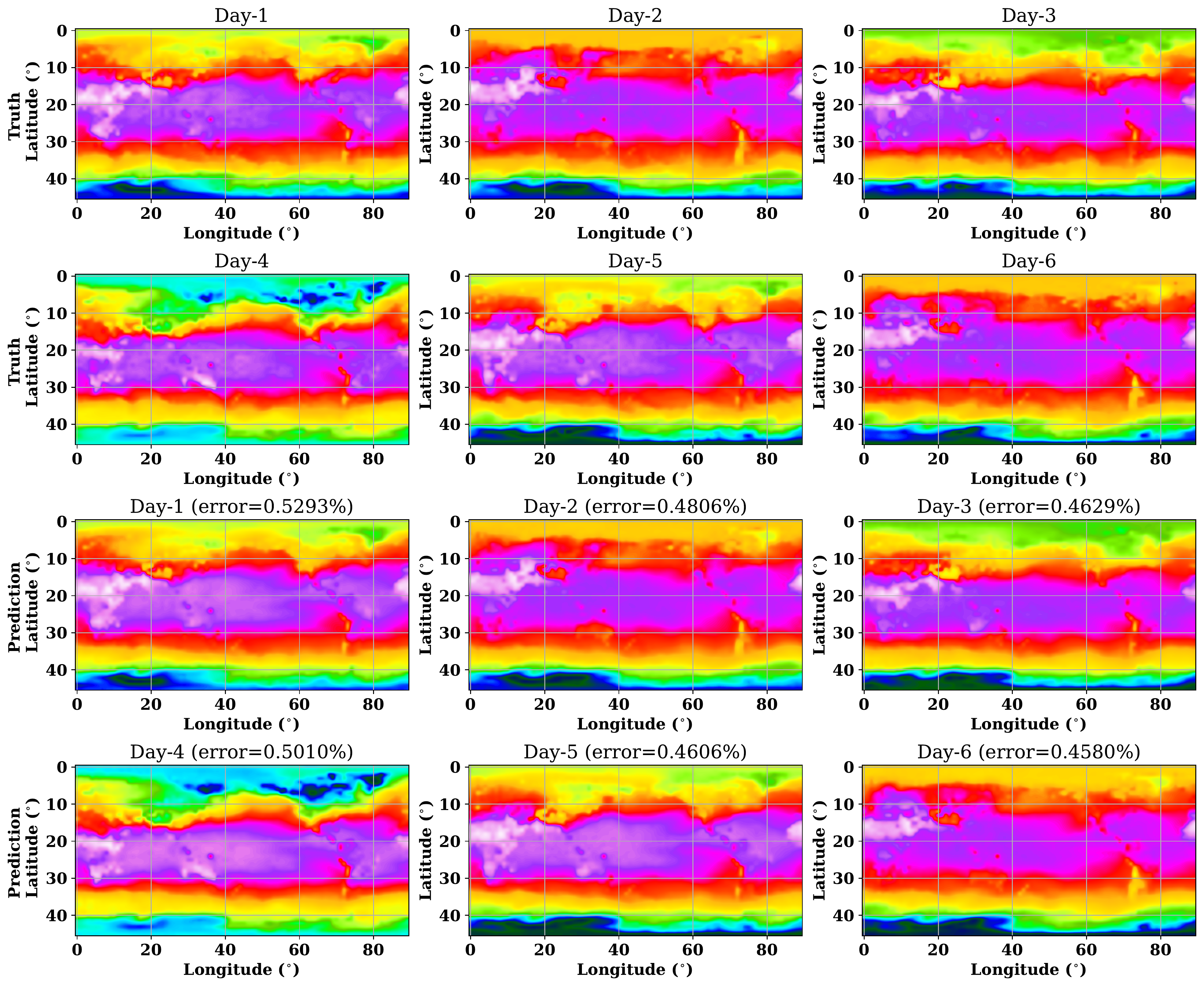}
	\caption{\textbf{WNO for forecast of the weekly 2m air temperature on a resolution of ${\mathbf{2^{\circ} \times 2^{\circ}}}$}. The predictions along with their corresponding errors for weekly forecast of 2m temperature at 12.00 UTC between $10^{th}$ to $15^{th}$ April are shown in the above figure. The errors are obtained with respect to the forecast made by Integrated Forecast System (IFS) of ECMWF. The predictions using proposed WNO are obtained with an error of the magnitude of $<1 \%$ that is a proof of the nearly accurate match with the actual forecast data. Once the training is done the predictions for the entire week are made within 3 seconds on the given GPU, which is evident of the computational advantage of the proposed WNO framework.}
	\label{fig_era5_time}
\end{figure}

The framework for short-term prediction of the 2m temperature is provided in Fig. \ref{fig_era5_climate}. The predictions are done at 12.00 Universal Time Coordinated (UTC) standard. For this purpose, the daily measurement of 2m temperature at 12.00 UTC time standards  from Jan 1st 2017 to Dec 31st 2021 are taken from ERA5 database. We first scaled down the data from $0.25^{\circ} \times 0.25^{\circ}$ to $2^{\circ} \times 2^{\circ}$ and arranged it into batches of 7, where the 7 denotes the number of days in a week. We then feed the dataset to the proposed WNO. Given the observations of the previous week the aim here is to predict the 2m temperatures at 12.00 UTC time standards for next 7 days. 
For this problem, a recurrent neural network (RNN) structure is formulated within the WNO. The WNO architecture consists of 4 WNO layer where db4 Daubechies wavelet with 2 level of decomposition is performed. A total number of 276 training samples, 6 testing samples and 500 epochs with batch size of 3 are used. The initial learning rate is kept at 0.001 with a step decay constant of 0.75 with step size 50. From previous 7 days database, the WNO performs the prediction for Day 8 and discards the Day 1 information from previous record by appending newly obtained Day 8 information to the previous database. This process is continued until the complete prediction for next 7 days are obtained. The prediction results are compared with the resolution-reduced results of Numerical Weather Prediction (NWP) system from ECMWF. The prediction results with their corresponding error are presented in Fig. \ref{fig_era5_time}. At the accuracy of $2^{\circ} \times 2^{\circ}$, the short-term weekly prediction results obtained using the proposed WNO is highly accurate. Close observations on the results reveal that the WNO has almost accurately captured the finest details present in the true results at $2^{\circ} \times 2^{\circ}$ resolution. Additionally, once trained, the predictions for entire week took only 3 second; this indicates computationally efficiency of the proposed WNO.

\begin{figure}[ht!]
	\centering
	\includegraphics[width=\textwidth]{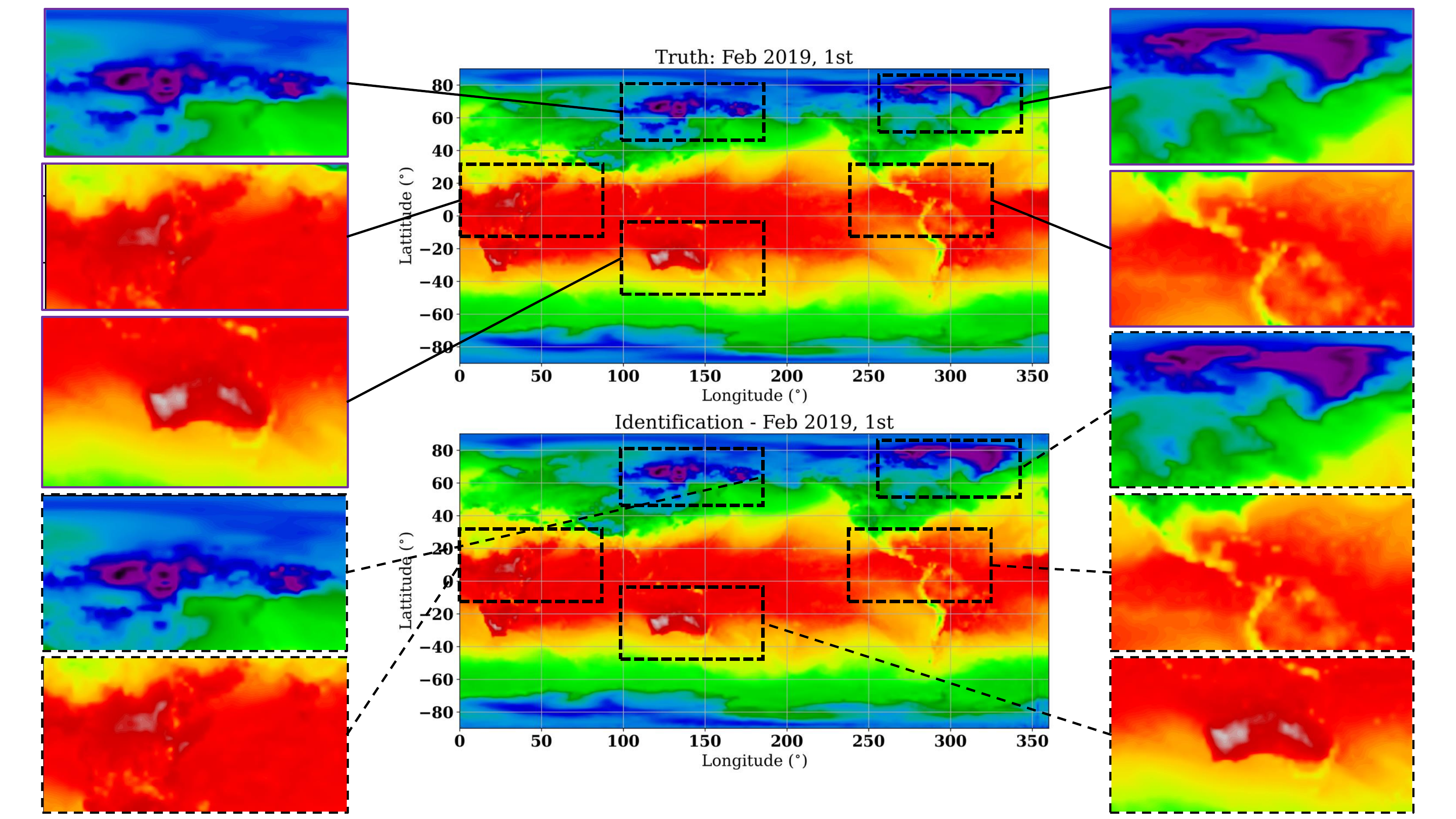}
	\caption{\textbf{WNO for forecast of the monthly averaged 2m air temperature on a resolution of ${\mathbf{2^{\circ} \times 2^{\circ}}}$}. The predictions for monthly averaged 2m temperature for 1st February of 2019 is performed. The datasets from Numerical Weather Prediction and proposed WNO are shown in the above figure. The predictions shows accurate match with the actual data with error of the magnitude of $\approx 0.1 \%$. On close observation it can be observed that, in addition to the actual temperature gradients the proposed WNO has also captured the finer details like white spots. Even in small resolutions the WNO has learned the small-scale features.}
	\label{fig_era5}
\end{figure}

Next, we focus on medium-range prediction of monthly mean 2m temperature at 12.00 UTC time standard. For this purpose, the monthly averaged datasets from 1979 to 2022 at 12.00 UTC standard are taken from ECMWF-ERA5 database. The problem is formulated as a simple 2D problem and the proposed WNO is trained from the previous datasets to perform the predictions for next monthly mean 2m temperature. For training and testing the total dataset is sliced into 480 and 40 samples and a total of 500 epochs with batch size 5 is used. The bd4 Daubechies wavelet with 5 level of decomposition are utilized. The initial learning rate is set as 0.001 which is later decayed with decay constant 0.75 at a step size of 50. The prediction results with corresponding errors are given in Fig. \ref{fig_era5}. 

From the presented results, it can be conjectured that the proposed WNO can successfully learn the operators of the various parameters of climate and atmosphere, thereby can be used for computationally faster and accurate predictions of arbitrary high-time scale parameters. We agree that the low-resolution based prediction are not enough to accurately capture the finest details of climate dynamics, however, in this example, we only take a small subset of all the possible applications and showcase the possibility for the proposed WNO for weather prediction. Given computational facilities, the WNO can also learn the climate dynamics at desired resolutions that will facilitate the predictions of small-scale features in the climate variables.

\section*{Discussion}
In this article, we proposed a new neural operator for learning of highly nonlinear operators that arise naturally in day-to-day scientific and engineering problems. Theoretically, the proposed WNO performs a local transformation on the input and then carries out a series of updates on the locally transformed inputs. The series of updates are formulated using a combination of nonlinear activation functions and kernel integral operators. The integral kernels are parameterized by neural network, where the parameterization of the integral kernel is done using a convolution type architecture; however, instead of performing the convolution directly in the physical space, we propose a wavelet kernel integral operator. Within wavelet kernel integral operator, 
the inputs are first transformed into the spectral domain using the wavelet and thereafter the convolutions are performed. Together with nonlinear activation functions and wavelet kernel integral updates, the proposed neural operator is able to represent any infinite-dimensional function space.
While parameterization of integral kernel, it exploits the spatial and frequency localisation property of the wavelets to learn the frequency of change in pixels over the spatial domain. The use of wavelet space for learning of integral kernel thus allows to track the finest patterns in the inputs; this allows the proposed WNO to learn the highly nonlinear operators even in complex geometric and boundary conditions. In terms of applications, we applied the proposed WNO to a variety of time-independent and time-dependent 1D and 2D PDE examples, exhibiting nonlinear spatial and spatio-temporal behaviours in flow, wave and phase-field problems.
The results obtained using the proposed WNO is compared with recently developed state-of-the art neural operators like DeepONet, FNO and MWT. The results suggests that the proposed WNO is able to learn the highly nonlinear differential operators very effectively even when complex geometry like triangular domain and notch was involved.
Quantitatively, for the six examples presented, the error in results obtained using WNO varies in the range 0.21\% -- 1.75\% with WNO outperforming the other operator learning approaches in four problems.

The proposed neural operator can learn the mapping between infinite-dimensional function spaces by combining the nonlinear activation functions with wavelet integral layers. We particularly exploits this fact and extends the application of proposed WNO beyond PDEs to satellite images. In a broader sense, the images and videos can also be contemplated as 2D and time dependent 2D problems, describing the motion or dynamics of an object/process. Thus, the proposed WNO can also be implemented on images and videos captured from cameras and satellite. As an possible application, we demonstrated the proposed WNO for prediction of weekly and monthly mean 2m air temperature of the Earth at a resolution of $2^{\circ} \times 2^{\circ}$. In the given resolution, the results are highly encouraging, the proposed framework produces highly accurate predictions for weekly and monthly forecast. 

It is worthwhile to note that the proposed WNO is highly efficient and hence, is ideally suited for scenarios where real-time predictions are necessary.
In this context, two interesting directions of future work includes application of WNO for development of digital twins and WNO based active vibration control. The weather prediction used-case presented in this paper can be regarded as a weak digital twin for Earth's climate. However, the framework in its current form can only predict the Earth's temperature (weekly and mean monthly). An extremely useful extension will be to develop a fully functional digital twin for Earth's climate that can also predict other climate-related parameters such as rainfall precipitation, storms, etc. Such digital twins can assist mankind in its fight against climate change and global warming. 


\section*{Methods: Wavelet Neural Operator for parametric partial differential equations}\label{sec:WNO}
Our aim is to perform the kernel integration in Eq. \eqref{eq:integral} in the neural network framework by parameterizing the kernel $k(a(x,y),x,y)$ with $\phi \in {\bm{\theta}}$. 
Before we construct the framework for parameterization in wavelet space, let us revisit the convolution integral. Let us drop the term $a(x,y)$ in Eq. \eqref{eq:integral} and modify the kernel as $k_{\phi}(x-y)$. By doing this we rewrite the Eq. \eqref{eq:integral} as,

\begin{equation}\label{eq:convolve}
    \left(K(\phi) * v_{j}\right)(x):=\int_{D \in \mathbb{R}^{d}} k \left(x- y; \phi \right) v_{j}(y) \mathrm{d} y; \quad x \in D, \quad j \in [1,l]
\end{equation}
which is consistent with the convolution operator. In this work, we propose to learn the kernel $k_{\phi}$ by parameterizing it in the wavelet domain. By learning kernel in wavelet domain we intend to perform the above convolution on the coefficients of wavelet decomposition rather than direct convolution in physical space. Let $\psi(x) \in \mathcal{L}^2(\mathbb{R})$ be an orthonormal mother wavelet that is localised both in the time and frequency domain. Let, $\mathcal{W}(\Gamma)$ and $\mathcal{W}^{-1}(\Gamma_W)$ be the forward and inverse wavelet transforms of an arbitrary function $\Gamma: D \mapsto \mathbb{R}^{d_{v}}$. Then the forward and inverse wavelet transforms  of the function $\Gamma$ with the scaling and translational parameters $\alpha \in \mathbb{R}^{+*}$ and $\beta \in \mathbb{R}$ are given by the following integral pairs,

\begin{equation}\label{eq:wavelet}
    \begin{aligned}
        (\mathcal{W} \Gamma)_{j}(\alpha, \beta) = \int_{D} \Gamma (x) \frac{1}{|\alpha|^{1 / 2}} \psi\left(\frac{x-\beta}{\alpha}\right) dx,    && \& &&    (\mathcal{W}^{-1} \Gamma_w)_{j}(x) = \frac{1}{C_{\psi}} \int_{0}^{\infty} \int_{D} (\Gamma_{w})_{j}(\alpha, \beta) \frac{1}{|\alpha|^{1 / 2}} \tilde{\psi}\left(\frac{x-\beta}{\alpha}\right) d\beta \frac{d\alpha}{\alpha^{2}},
    \end{aligned}
\end{equation}
where $(\Gamma_{w})_{j}(\alpha, \beta) = (\mathcal{W} \Gamma)_{j}(\alpha, \beta)$ $\psi((x-\beta) \alpha) \in \mathcal{L}^2(\mathbb{R})$ is a scaled and translation of mother wavelet, often called as the daughter wavelet. By scaling and shifting, the desired wavelets can be obtained from the mother wavelet. The term $C_{\psi}$ is called as the admissible constant whose range is given as $0 < C_{\psi} < \infty$. The expression for $C_{\psi}$ is given as,

\begin{equation}
    C_{\psi}= 2\pi \int_{D} \frac{{|\psi (\omega)|}^2}{|\omega|} d \omega,
\end{equation}
where $\psi (\omega)$ denotes the Fourier transform of $\psi (x)$. Applying the convolution theorem, we therefore obtain that,

\begin{equation}\label{eq:conv_pre}
    \begin{aligned}
        \left(K(a ; \phi) * v_{j}\right)(x) = \mathcal{W}^{-1}\left( \mathcal{W}(k_{\phi}) \cdot \mathcal{W} (v_{j})\right)(x); && x \in D.
    \end{aligned}
\end{equation}
Since we propose to parameterize the kernel $k_{\phi}$ directly in wavelet space instead of performing wavelet transform of $k_{\phi}$, we represent $R_{\phi} = \mathcal{W}(k_{\phi})$, where $R_{\phi}$ represents the wavelet transform of the kernel function $k:D \mapsto \mathbb{R}^{d_{v} \times d_{v}}$. With the above information, we define the wavelet neural operator as,

\begin{equation}\label{eq:conv_final}
    \begin{aligned}
        \left(\mathcal{K}(\phi) * v_{j}\right)(x)=\mathcal{W}^{-1}\left(R_{\phi} \cdot \mathcal{W}( v_{j})\right)(x); && x \in D.
    \end{aligned}
\end{equation}
One issue with continuous wavelet transform (CWT) is that it estimates the wavelet coefficients at all possible scales i.e., they contain information along an infinite number of wavelets, which may be redundant and computationally expensive. However, the same equivalent accuracy with computationally speedup can be achieved by performing the discrete wavelet transform (DWT). In DWT, we modify the mother wavelet to calculate the wavelet coefficients at scales with powers of two. In this case, the wavelet $\phi(x)$ is defined as, 

\begin{equation}
    \psi_{m, \tau}(x) = \frac{1}{\sqrt{2^{m}}} \psi\left(\frac{x - \tau 2^{m}}{2^{m}}\right)
\end{equation}
where the parameter $m \in \mathbb{Z}^{+}$ and $\tau \in \mathbb{Z}^{+*}$ are the scaling and translational parameters. The admissible constant in DWT is given by,

\begin{equation}
    C_{\psi}= \int_{0}^{\infty} \frac{{|\psi (\omega)|}^2}{|\omega|} d \omega,
\end{equation}
The forward DWT is given as,

\begin{equation}
    \begin{aligned}
        (\mathcal{W} \Gamma)_{j}(m, \tau)=\frac{1}{\sqrt{2^{m}}} \int_{D} \Gamma (x) \psi\left(\frac{x-\tau 2^{m}}{2^{m}}\right) dx.
    \end{aligned}
\end{equation}
By fixing the scale parameter $m$ at a particular integer and shifting the $\tau$, only the DWT coefficients at level $m$ can be obtained. Though there are many different flavours of DWT, we particularly chose the filterbank implantations of DWT. The filterbank implementation decomposes the signal into detail and approximation coefficients by passing the signal through a low-pass and a high-pass filter. If $r(n)$ and $s(n)$ denote the low-pass and high-pass filter, respectively, then two convolutions of the form $z_{low}(n) = (x*r)(n) \downarrow 2$ and $z_{high}(n) = (z*s)(n) \downarrow 2$ are executed, where $n$ is the number of discretization points. While the detail coefficients $z_{high}(n)$ are retained, the approximation coefficients are recursively filtered by passing it through the low-pass and high-pass filters until the total number of decomposition levels are exhausted. At each level, the length of the signal is halved by a factor 2 due to the conjugate symmetry. At the end of entire operation, the length of the support width of wavelet coefficients is given by the formula: $\zeta = n_{s}/2^m + (2n_{w}-2)$, where, $n_{s}$ is the length of the signal, $m$ is the decomposition level in DWT and $n_{w}$ is the length of the vanishing moment of the undertaken wavelet.

In the DWT, the coefficients in smaller scales such as $2^0$ and $2^1$ are usually associated with high frequencies; thus, they are not the canonical choice for kernel parameterization. On the other hand, the high scale wavelet coefficients generally corresponds to low frequency information thus containing the most important features of the input signal. We thus, choose to parameterize $R_{\phi}$ on higher scales of discrete wavelet transform, possibly the last level. For this purpose, a finite-dimensional parameterization space is obtained by keeping the information from only the highest scale wavelet coefficients. For $n \in D$, we have $v_{t} \in \mathbb{R}^{n \times d_{v}}$, $\mathcal{W}\left(v_{t}\right) \in \mathbb{R}^{n \times d_{v}}$ and $R_{\phi}(m) \in \mathbb{R}^{d_{v} \times d_{v}}$. Since we want to parameterize $R_{\phi}$ only on the highest level of decomposition, we have $\mathcal{W}\left(v_{t}\right) \in \mathbb{R}^{n/2^m \times d_{v}}$ in a $m$-level of DWT. In general, the length of wavelet coefficients are also influenced by the number of vanishing moments of the orthogonal mother wavelet. Thus, it is more appropriate to write that $\mathcal{W}\left(v_{t}\right) \in \mathbb{R}^{\zeta \times d_{v}}$. The direct parameterization of $R_{\phi}$ in the wavelet space is therefore carried out by integrating out the parameter $\phi$ as convolution of $\left(\zeta \times d_{v} \times d_{v} \right)$-valued tensor. Rephrasing Eq. \eqref{eq:conv_final}, we can write the convolution of weight tensor $R \in \mathbb{R}^{\zeta \times d_{v} \times d_{v}}$ and $\mathcal{W}\left(v_{t}\right) \in \mathbb{R}^{\zeta \times d_{v}}$ as,

\begin{equation}
    \begin{aligned}
        \left(R \cdot \mathcal{W} (v_{j})\right)_{I_1, I_2} (x) = \sum_{I_{3}=1}^{d_{v}} R_{I_1, I_2, I_3} \mathcal{W} (v_{j})_{I_1, I_3}; && I_1 \in [1, \zeta], && I_{2}, I_{3} \in d_{v}.
    \end{aligned}
\end{equation}

\subsection*{Relation with Fourier Neural Operator (FNO)}
The proposed WNO is closely related to FNO; infact, we will show that the FNO is a special case of the proposed WNO. To understand the relation, let us consider the generalised wavelet transform that is given as,

\begin{equation}\label{eq:gen_wav}
\mathcal{W}(\alpha, \beta) = \int_{D} \Gamma(x) \psi_{\alpha, \beta}(x) dx,
\end{equation}
where $\psi_{\alpha,\beta}(x) \in \mathcal{L}^2(\mathbb{R})$ is the orthonormal wavelet and $\alpha$, $\beta$, $\Gamma(x)$ denote the same as above. In wavelet transform, the aim is to transforms the continuous function $\Gamma (x)$ into a continuous function of scale and translational variables $\alpha$ and $\beta$. The transformation is generally done by projecting $\Gamma (x)$ into different wavelet coefficients using the following form of $\psi_{\alpha,\beta}(x)$, 

\begin{equation}
\psi_{\alpha, \beta}(x) = \frac{1}{\sqrt{\alpha}} \psi\left(\frac{x-\beta}{\alpha}\right)
\end{equation}
where $\psi_{\alpha, \beta}(x)$ will decide the nature of waveforms. Based on the requirements, if one chose to project the function $\Gamma (x)$ in terms of its frequency components, the translation parameter can be dropped and the orthonormal basis wavelet can be replaced using the triangular/harmonic basis functions. If one chose to consider only the sine and cosine waves as the orthonormal wavelets i.e., to replace the orthonormal wavelet $\psi_{\alpha,\beta}(x)$ by $e^{-2 i \pi\langle x, \omega\rangle}$, Eq. \eqref{eq:gen_wav} will result in a standard Fourier transform. If we denote $\mathcal{F}(\omega)$ as the Fourier transform and $\mathcal{F}^{-1}(x)$ to be its inverse operation, then for the function $\Gamma (x)$, we have,

\begin{equation}
    \begin{aligned}
        (\mathcal{F} \Gamma)_{j}(\omega)=\int_{D} f_{j}(x) e^{-2 i \pi\langle x, \omega \rangle} dx, && \& && \quad\left(\mathcal{F}^{-1} \Gamma \right)_{j}(x)=\int_{D} f_{j}(\omega) e^{2 i \pi\langle x, \omega \rangle} \mathrm{d} \omega; &&  j=1, \ldots, d_{v}
    \end{aligned}
\end{equation}
Since the Fourier transform can translate between the convolution of multiple functions, Eq. \eqref{eq:conv_pre} can be written in Fourier space as,

\begin{equation}
    \begin{aligned}
        \left(\mathcal{\omega}(a ; \phi) v_{j}\right)(x)=\mathcal{F}^{-1}\left(\mathcal{F}(k_{\phi}) \cdot \mathcal{F}(v_{j})\right)(x);  && x \in D.
    \end{aligned}
\end{equation}
The kernel $k_{\phi}$ therefore can be directly parameterized in Fourier space by choosing a sufficient number of Fourier modes $\omega_{\max} \in D$.

\subsection*{Note on the complexity of the wavelet integral layer}
Without loss of generality, we restrict the complexity analysis to a 1D problem. Recall, $n$ is the number of discretization points in the domain $D$. In the discrete wavelet decomposing, the input signals are decomposed simultaneously using a low-pass filter $r(n)$ and a high-pass filter $s(n)$. In case of Daubechies wavelets, the filters $r(n)$ and $s(n)$ have a constant length, each of which performs the convolutions $(x*r)(n)$ and $(x*s)(n)$, thereby has the complexity $\mathcal{O}(n)$.
In the single tree format, the discrete wavelet decomposition uses these two filters for recursive multi-level decomposition of the output of the branch convolved with $r(n)$. This recursive filterbank implementation has the overall complexity of $\mathcal{O}(n)$. Decomposition of the input using wavelet with a $M$ level results in a signal of length $n/2^{M}$. Convolution of decomposed coefficients and weight tensor inside the wavelet integral layer takes $\mathcal{O}(n/2^{M})$ time.
The convolution in the CNN with kernel size takes $\mathcal{O}(n)$ time. Therefore, the majority of the computational demand arises from the convolution in wavelet integral. In general, there is a trade between the number of decomposition and the integral kernel size and their relation can be represented as a reciprocal relation, i.e., more the decomposition level less the integral kernel size and vice-versa. However, we observed that the level of decomposition play a significant role in speed-up of the proposed neural operator as compared to the size of integral kernel. For discrete wavelet decomposition we make use of the pytorch-wavelet and pywavelet libraries, more details can be found in Ref. \cite{lee2019pywavelets,cotter2020uses}.

\section*{Acknowledgements}
T. Tripura acknowledges the financial support received from the Ministry of Human Resource Development (MHRD), India in form of the Prime Minister's Research Fellows (PMRF) scholarship. S. Chakraborty acknowledges the financial support received from Science and Engineering Research Board (SERB) via grant no. SRG/2021/000467 and seed grant received from IIT Delhi.

\section*{Data availability}
On acceptance all the used datasets in this study will be made public on GitHub by the corresponding author.

\section*{Code availability}
On acceptance all the source codes to reproduce the results in this study will be made available to public on GitHub by the corresponding author.

\end{document}